\documentstyle[11pt,subeqn]{article}

\newcommand{\sect}[1]{\setcounter{equation}{0}\section{#1}}
\renewcommand{\theequation}{\thesection.\arabic{equation}}
\newcommand{\EQ}{\begin{equation}}
\newcommand{\EN}{\end{equation}}
\newcommand{\bea}{\begin{eqnarray}}
\newcommand{\ena}{\end{eqnarray}}

\newcommand{\shalf}{\frac{1}{2}}
\begin{document}

\topmargin -15mm
\oddsidemargin 5mm

\renewcommand{\Im}{{\rm Im}\,}
\newcommand{\NP}[1]{Nucl.\ Phys.\ {\bf #1}}
\newcommand{\PL}[1]{Phys.\ Lett.\ {\bf #1}}
\newcommand{\NC}[1]{Nuovo Cimento {\bf #1}}
\newcommand{\CMP}[1]{Comm.\ Math.\ Phys.\ {\bf #1}}
\newcommand{\PR}[1]{Phys.\ Rev.\ {\bf #1}}
\newcommand{\PRL}[1]{Phys.\ Rev.\ Lett.\ {\bf #1}}
\newcommand{\MPL}[1]{Mod.\ Phys.\ Lett.\ {\bf #1}}
\newcommand{\IJMP}[1]{Int.\ Jour.\ of\ Mod.\ Phys.\ {\bf #1}}
\renewcommand{\thefootnote}{\fnsymbol{footnote}}

\newpage
\setcounter{page}{1}
\rightline{DFTT 33/94}
\rightline{DFT-US 2/94}
\vspace{1.4cm}
\begin{center}
{\Large {\bf{ANYONS AND
DEFORMED LIE ALGEBRAS}}\footnote{To appear in the
Proceedings of the International
School of Physics ``E. Fermi'',
Course CXXVII {\it Quantum Groups and Their
Applications in Physics} (28 June - 8 July 1994).} }
\vspace{1.5cm}

{\Large M. Frau $^{a}$, A. Lerda $^{b}$ and S. Sciuto $^{a}$}
\vspace{0.4cm} \\
$^{a}$ {\em Dipartimento di Fisica Teorica, Universit\'a di Torino} \\
{\em and I.N.F.N., Sezione di Torino, Italy}
\vspace{0.2cm} \\
$^{b}$ {\em Dipartimento di Fisica Teorica, Universit\'a di Salerno} \\
{\em and I.N.F.N., Sezione di Napoli, Italy} \\
\end{center}
\vspace{3cm}
\centerline{\large \bf Abstract}
We discuss the connection between anyons
(particles with fractional statistics)
and deformed Lie algebras (quantum groups). After a brief review
of the main properties of anyons, we present
the details of the anyonic realization of all deformed classical Lie
algebras in terms of anyonic oscillators. The
deformation parameter
of the quantum groups is directly related to the statistics
parameter of the anyons. Such a realization is a direct
generalization
of the Schwinger construction in terms of fermions
and is based on a sort of
bosonization formula which yields the generators of the deformed
algebra in terms of the undeformed ones. The entire procedure is
well defined on two-dimensional lattices, but it can be consistently
reduced also to one-dimensional chains.
\newpage
\renewcommand{\thefootnote}{\arabic{footnote}}
\setcounter{footnote}{0}
\vspace{1cm}
%
%

\sect{Introduction}
\vspace{0.7cm}

In this contribution we discuss the connection
between anyons and quantum groups that we originally
discovered in \cite{LS} and later extended in \cite{FMS}
in collaboration with M.A. R.-Monteiro (see also \cite{CM}).
Anyons and quantum groups seem apparently two very distinct subjects,
but as we shall see, they share one common important
property: both of them are deeply related
to the braid group.

Anyons are particles with {\it any} statistics
\cite{LM,W1} that interpolate
between bosons and fermions of which they can be considered, in some
sense, a deformation (for reviews see for instance \cite{W2,L}).
Anyons exist only in two dimensions because in this
case the configuration space of collections of identical particles
has some special topological properties allowing arbitrary
statistics, which do not
exist in three or more dimensions.
Specifically the configuration space of identical
particles is infinitely connected in two dimensions, but
it is only doubly connected in three or more dimensions.
In this  case only two statistics are possible (the bosonic
and fermionic ones) whereas in two
dimensions there are infinitely
many possibilities.
Anyons are not objects
of only pure mathematical interest; on the contrary they play
an important role in certain systems of the real world.
Since this has at least three space dimensions,
it is clear that anyons cannot be real particles.
However, there
exist some (condensed matter) systems
that can be regarded effectively as two-dimensional and in which
the localized quasi-particle excitations obey the rules
of the two-dimensional world. It is these quasi-particles
that may be anyons and have fractional statistics. The most
notable example of this behavior
is provided by the systems that exhibit
the fractional quantum Hall effect \cite{PG} in
which the localized
excitations carry fractional charge, fractional spin and
fractional statistics, and are therefore anyons \cite{LH}.

Since the first paper on the subject \cite{LM}, it became clear that
anyons are deeply connected to the braid group of
which they are abelian representations, just like
bosons and fermions are abelian representations
of the permutation group. In fact, when one exchanges two identical
anyons, it is not enough to compare their final configuration
with the initial one,
but instead it is necessary to specify also the way in which
the two particles are exchanged, {\it i.e.} the way in which they
braid aroud each other.

The braid group plays a crucial role also in the theory of
quantum groups (or more properly of quantum universal
enveloping algebras) of which
it is the centralizer \cite{D,J,Wo,FRT}.
Quantum groups are deformations of ordinary Lie algebras
which have recently found interesting applications in several
areas of physics, like in the theory of exactly solvable
models \cite{ESM,FRT}, in conformal field theories \cite{CFT} or
in condensed matter theory \cite{CMa}.
It has been conjectured also that
quantum groups might be the characteristic symmetry structures
of anyon systems, even though explicit realizations
of this fact are still missing. However, the
fundamental role played
by the braid group both in the theory of quantum groups
and in the theory of anyons suggests that at least a direct
relation between the two subjects should exist.
It is well known that bosonic or fermionic oscillators,
characterized by commutative
or anti-commutative Heisenberg algebras,
can be combined \`a la
Schwinger \cite{S} to construct non-abelian Lie algebras $G$
with the permutation group as centralizer.
Similarly one can think of
using anyonic oscillators with braid group properties and
deformed commutation relations to build
non-abelian algebras with
the braid group as centralizer, {\it i.e.} to construct non-abelian
quantum groups ${\cal U}_q(G)$ from anyons.
Here we elaborate on this idea
and show that actually all deformed classical Lie algebras
admit a simple anyonic realization.
In particular, we will show \cite{LS,FMS,CM} that using anyons of
statistics $\nu$ it is possible to realize the quantum universal
enveloping
algebras ${\cal U}_q(A_r)$, ${\cal U}_q(B_r)$, ${\cal U}_q(C_r)$
and ${\cal U}_q(D_r)$ with $q=\exp({\rm i}\,\nu\,\pi)$.

A unified treatment for all these cases
is provided by a sort of
bosonization formula which expresses
the generators
of the deformed algebras in terms of the undeformed ones.
Such a bosonization resembles the one typical of
quantum field theories in $1+1$ dimensions
\cite{CoM} where bosons and
fermions can be related to each other by means of the so-called
disorder operators, and is similar also
to the anyonization formula characteristic of
quantum field theories in $2+1$ dimensions
\cite{W1,JP}.
However, we would like to stress that our bosonization formula is
different from the standard relation between the generators of
quantum and classical algebras.
In fact, our expression is strictly
two-dimensional and non-local
since it involves anyonic operators defined
on a two-dimensional lattice. Thus it
cannot be extended to higher dimensions
in a straightforward way.
However we remark that
anyons can consistently be defined also on one dimensional chains;
in such a case they become local
objects and their braiding
properties are dictated by their natural ordering on the line.
Consequently, our construction
can be used equally well for one dimensional chains.

We organize this contribution as follows. In Section 2 we
present a brief introduction to anyons and discuss how they
can be realized from fermions by means
of a generalized Jordan-Wigner
transformation. In Section 3 we review the
basic properties of deformed Lie
algebras that will be needed later.
In Section 4 and 5 we present the details of our anyonic
realization of deformed classical Lie algebras and finally in
the last section we present some conclusive remarks.

%
%
\sect{Anyonic Statistics in Two Dimensions}
\vspace{0.7cm}

In first quantization the notion of statistics is associated to the
properties
of the wave functions describing systems of identical particles under
the exchange of any two of these. If the wave function
is totally symmetric, it describes bosons; if instead it is
totally antisymmetric,
it describes fermions. However, in two space dimensions there exist more
possibilities and the wave function for a system of
identical particles is in general neither
symmetric nor antisymmetric under permutations, but
acquires a phase which generalizes the plus sign typical of bosons
and the minus sign typical of fermions. In this general case
one says that the
wave function describes {\it anyons}, particles of arbitrary statistics.
For example, a typical wave function for anyons of statistics $\nu$
is
\EQ
\psi(z_1,z_1^*;...;z_N,z_N^*) = \prod_{I<J}
\left(z_I-z_J\right)^\nu\,f(z_1,z_1^*;...;z_N,z_N^*)\ \ ,
\label{2.0}
\EN
where $z_I=x_I+{\rm i}\,y_I$ is the position of the $I$-th particle
in complex notation and $f$ is a single-valued function symmetric
under all permutations.
When particle $I$ is exchanged with particle $J$,
$\psi$ acquires a phase ${\rm e}^{{\rm i}\,\pi\,\nu}$
or ${\rm e}^{-{\rm i}\,\pi\,\nu}$ depending on whether the exchange is
done by rotating $I$ around $J$
clockwise or counterclockwise. Thus, if $\nu\not= 0,1$ (mod 2) it is
of fundamental importance to specify not only the
permutation but also the orientation of the exchange, {\it i.e.}
the braiding. This is the essential
feature that distinguishes anyons from bosons
and fermions.

In second quantization the notion of statistics is
usually associated to the algebra that the
particle creation and annihilation operators satisfy. In fact, bosonic
operators close canonical commutation relations, whilst
fermionic operators
close canonical anticommutation relations. Therefore, in two dimensions
where statistics is arbitrary there should exist also anyonic
creation and annihilation operators which close braiding
relations among themselves. As we will see in
the following, these anyonic operators
do indeed exist and can be simply constructed from fermionic
(or bosonic) operators using a generalized Jordan-Wigner
transformation.

Let us now briefly review \cite{L} the main features of anyonic
statistics, both
from the first and the second quantization standpoints.
We start by considering a system of $N$
indistinguishable hard-core particles
moving in ${\bf R}^d$. The configuration space for this system is
\EQ
M_N^d = {{\left({\bf R}^d\right)^N -\Delta}\over S_N}\ \ ,
\label{2.1}
\EN
where $\Delta$ is the set of all points in $\left({\bf R}^d\right)^N$
with at least two equal coordinates and $S_N$ is the permutation group of
$N$ objects. In (\ref{2.1}) $\Delta$
is removed from $\left({\bf R}^d\right)^N$
because of the hard-core condition which prevents any two particles
from occupying the same position, and $S_N$ is moded out because
any two configurations which simply differ by a
permutation must be identified
since the particles are indistinguishable.

The topology of $M_N^d$ is radically different depending on whether
$d>2$ or $d=2$. For example, if $d>2$ the fundamental group of $M_N^d$
is the permutation group,
{\it i.e.}
\EQ
\pi_1\left(M_N^d\right) = S_N\ \ .
\label{2.2}
\EN
On the contrary, if $d=2$ we have
\EQ
\pi_1\left(M_N^2\right) = B_N\ \ ,
\label{2.3}
\EN
where $B_N$ is the braid group of $N$ objects.
It is precisely this topological property that allows the
existence of anyonic statistics in $d=2$.

To see this, let us
consider the probability amplitude that our system evolves from
a certain configuration $q\in M_N^d$ at time $t$ to the same
configuration at a later time $t'$. Let us denote this amplitude
by $K(q;t,t')$.
In the path-integral formulation of quantum mechanics, $K(q;t,t')$
is represented by a sum over {\it all} loops in $M_N^d$ starting
from $q$ at time $t$ and arriving at $q$ at time $t'$. This sum
can be organized as a sum over
homotopy classes $\alpha \in \pi_1\left(M_N^d\right)$
and a sum over all elements $q_\alpha$
of each class. Within each homotopy
class, all loops are weighted with the exponential of the action, but
different classes in general can have different weights \footnote{For
bosons all classes have the same weight, but for
fermions each class is weighted with a plus
or a minus sign depending on the
parity of the corresponding permutation.}. In fact,
denoting by ${\cal L}$ the Lagrangian of the system, one has
\EQ
K(q;t,t') = \sum_{\alpha}
\chi(\alpha)
\int{\cal D}q_\alpha
{}~ {\rm e}^{\,{\scriptstyle{\rm i}\over\scriptstyle{\hbar}}
\int_{t}^{t'}d\tau ~{\cal L}}
\label{2.4}
\EN
where $\chi(\alpha)$ is a complex number representing the
weight with which the class $\alpha$
contributes to the path-integral.
This complex number cannot be totally arbitrary; indeed
if we want to mantain the standard rules for combining
probability amplitudes, it is necessary that
\EQ
\chi(\alpha_1)\,\chi(\alpha_2) = \chi(\alpha_1\,\alpha_2)
\label{2.5}
\EN
for any $\alpha_1$ and $\alpha_2$. Thus, $\chi(\alpha)$ must be
a one-dimensional representation of the fundamental group
of the configuration space of the system, which for $d=2$ is the
braid group (see (\ref{2.3}))
\footnote{In principle one could also allow non-abelian
representations of
the fundamental group. Such possibility, which we do not discuss here,
leads to parastatistics for the permutation group ($d>2$) and to plectons
for the braid group ($d=2$).}.
Let us recall that the braid group of $N$ objects $B_N$ is
an infinite group generated by $N-1$
elements $\{\sigma_1,...,\sigma_{N-1}\}$
which satisfy
\EQ
\sigma_I\,\sigma_{I+1}\,\sigma_I=\sigma_{I+1}\,\sigma_I\,\sigma_{I+1}
\label{2.6}
\EN
for $I=1,...,N-2$, and
\EQ
\sigma_I\,\sigma_J = \sigma_J\,\sigma_I
\label{2.7}
\EN
for $|I-J|\geq 2$. The generator $\sigma_I$ simply represents
the exchange of particle $I$ and particle $I+1$ with a definite
orientation (say counterclockwise). Any word constructed with the
$\sigma_I$'s and their inverses, modulo the relations
(\ref{2.6}-\ref{2.7}), is an element of $B_N$. Notice that
$\sigma_I^2 \not= 1$. If  we impose the further condition
$\sigma_I^2=1$ for all $I$, then the braid group reduces
to the permutation group $S_N$. The one-dimensional unitary
representations
of $B_N$ are simply given by
\EQ
\chi(\sigma_I) = {\rm e}^{-{\rm i}\,\nu\,\pi}~~~~~\forall~ I\ \ ,
\label{2.8}
\EN
where $\nu$ is an arbitrary real parameter labelling the representation.
Clearly we can always restrict $\nu$ in the interval $[0,2)$.
For an arbitrary braiding $\alpha \in B_N$,
(\ref{2.8}) is generalized as follows
\EQ
\chi(\alpha) = {\rm e}^{-{\rm i}\,\nu\,\pi\,P_\alpha}\ \ ,
\label{2.9}
\EN
where $P_\alpha$ is the difference between the number of counterclockwise
exchanges and the number of clockwise exchanges that occur in
$\alpha$. When we insert (\ref{2.9}) into (\ref{2.4}) we obtain
\EQ
K(q;t,t') = \sum_{\alpha \in B_N}
{\rm e}^{-{\rm i}\,\nu\,\pi\,P_\alpha}
\int {\cal D}q_\alpha
{}~ {\rm e}^{\,{\scriptstyle{\rm i}\over\scriptstyle{\hbar}}
\int_{t}^{t'}d\tau ~{\cal L}}
\label{2.10}
\EN
which represents the first-quantized propagator for anyons
of statistics $\nu$. If $\nu=0$, we describe bosons, since all
homotopy classes of loops enter the path-integral with equal weight.
If $\nu=1$, we describe fermions since each homotopy class
$\alpha$ is weighted with a plus sign or a minus sign depending on the
parity of $P_\alpha$. If $\nu\not=0,1$, the weights of the homotopy
classes are generic phases and thus we describe
anyons of statistics $\nu$.

Let us now observe that (\ref {2.9}) can be written also as follows
\EQ
\chi(\alpha) = {\rm e}^{-{\rm i}\,\nu
\sum\limits_{I<J}\big[
\Theta_{IJ}^{(\alpha)}(t')-\Theta_{IJ}^{(\alpha)}(t)\big]}~=~
{\rm e}^{-{\rm i}\,\nu
\sum\limits_{I<J}\int_t^{t'}d\tau \ {{d}\over{d\tau}}
\Theta_{IJ}^{(\alpha)}(\tau)}\ \ ,
\label{2.11}
\EN
where $\Theta_{IJ}^{(\alpha)}(t)$ is the winding angle
of particle $I$ with
respect to particle $J$ measured along the braiding $\alpha$ at time $t$.
The function $\Theta_{IJ}^{(\alpha)}(t)$ is very
complicated, but its explicit expression is not needed in the
following. Inserting (\ref{2.11}) into (\ref{2.4}), we get
\EQ
K(q;t,t') = \sum_{\alpha \in B_N}
\int {\cal D}q_\alpha
{}~ {\rm e}^{\,{\scriptstyle{\rm i}\over\scriptstyle{\hbar}}
\int_{t}^{t'}d\tau ~\left[{\cal L}-\hbar\,
\nu\sum\limits_{I<J}{{d}\over{d\tau}}
\Theta_{IJ}^{(\alpha)}(\tau)\right]}
\label{2.12}
\EN
which has an interesting interpretation. In fact
$K(q;t,t')$ is decomposed into subamplitudes
which are weighted equally as
if we were describing bosons with a Lagrangian
\EQ
{\cal L}_B={\cal L}-\hbar\,
\nu\sum\limits_{I<J}{{d}\over{d\tau}}
\Theta_{IJ}^{(\alpha)}(\tau)\ \ .
\label{2.13}
\EN
Furthermore, if we set
\EQ
\nu = 1+{\nu}'\ \ ,
\label{2.14}
\EN
we obtain
\EQ
K(q;t,t') = \sum_{\alpha \in B_N} (-1)^{P_\alpha}
\int {\cal D}q_\alpha
{}~ {\rm e}^{\,{\scriptstyle{\rm i}\over\scriptstyle{\hbar}}
\int_{t}^{t'}d\tau ~\left[{\cal L}-\hbar\,
{\nu}'\sum\limits_{I<J}{{d}\over{d\tau}}
\Theta_{IJ}^{(\alpha)}(\tau)\right]}
\label{2.15}
\EN
which too has an interesting interpretation. In fact, $K(q;t,t')$ is
decomposed into subamplitudes which contribute to the
path-integral with alternating signs as if we
were describing fermions with a Lagrangian
\EQ
{\cal L}_F={\cal L}-\hbar\,
{\nu}'\sum\limits_{I<J}{{d}\over{d\tau}}
\Theta_{IJ}^{(\alpha)}(\tau)\ \ .
\label{2.16}
\EN
Eqs. (\ref{2.13}) and (\ref{2.16}) mean that we can trade anyonic
statistics for some kind of ``fictitious'' force and describe anyons
as ordinary bosons or fermions with the modified Lagrangians
${\cal L}_B$ or ${\cal L}_F$. Notice that the new terms that appear in
in ${\cal L}_B$ or ${\cal L}_F$ are topological quantities ({\it i.e.}
total derivatives) which do not modify the local
dynamical properties of the particles like for example
their equations of motions, but which modify significantly their global
dynamical properties like their statistics. That is why these terms
are called statistical interaction terms.

We now present an explicit realization of these statistical
interactions by means
of Chern-Simons fields \cite{CS,W1}. Let us consider a system of $N$
non-relativistic point particles moving on a plane,
with mass $m$ and charge $e$,
whose coordinates ${\vec r}_I(t)$ and velocities ${\vec v}_I(t)$
serve as dynamical variables ($I=1,...,N$).
For definiteness we take these particles to be spinless
fermions which automatically satisfy the hard-core constraint because
of the Pauli exclusion principle. Their dynamics
is governed by the action
\[
S= \int \!dt \,{\cal L}\ \ ,
\]
where
\EQ
{\cal L} = \sum_{I=1}^N {1\over 2}\,m\,{\vec v}_I(t)^2
- V\left({\vec r}_1(t),...,{\vec r}_N(t)\right)\ \ ,
\label{2.17}
\EN
and $V\left({\vec r}_1(t),...,{\vec r}_N(t)\right)$ is the potential.
In this system there exists a conserved
current $j^\alpha(\vec{x},t)$, whose
components are the matter density
\EQ
j^0(\vec{x},t) \equiv \rho(\vec{x},t)
= \sum_{I=1}^N \delta({\vec x} -
{\vec r}_I(t))
\label{2.18}
\EN
and the current density
\EQ
{\vec j}(\vec{x},t) = \sum_{I=1}^N {\vec v}_I(t)\,\delta({\vec x} -
{\vec r}_I(t))
\label{2.19}
\EN
which satisfy the continuity equation
$~\partial_t\rho+{\vec \nabla}\cdot{\vec j}=0~$.

Now let us suppose that $j^\alpha$ be coupled to an abelian
gauge field $A_\alpha$ in the standard minimal way, {\it i.e.}
\EQ
S_{\rm int} = -e\,\int d^3x ~j^\alpha(x)\,A_\alpha(x)\ \ ,
\label{2.20}
\EN
and that the dynamics of $A_\alpha$ be governed by the
Chern-Simons action \cite{CS}
\EQ
S_{\rm CS} = {\kappa\over 2}\int d^3x ~\epsilon^{\alpha\beta\gamma}
\,A_\alpha(x)\,\partial_\beta\,A_\gamma(x)\ \ ,
\label{2.21}
\EN
where $\kappa$ is a coupling constant. (Here and in the following,
we set $c=\hbar=1$, assume
that the space-time indices are contracted, raised and lowered with
the Minkowski
metric with signature ${+,-,-}$, and denote
$({\vec x},t)$ simply by $x$ wherever
not ambiguous.)
The total action for
this interacting system is therefore
\EQ
S_{\rm tot} = S + S_{\rm int} + S_{\rm CS}
\label{2.22}
\EN
and is invariant under standard abelian gauge transformations.

What we want to show now is that the Chern-Simons field
effectively introduces a statistical
interaction among the particles,
and in particular that the
Lagrangian corresponding to $S_{\rm tot}$ is of the type
(\ref{2.16}). To see this, let us first
observe that varying $S_{\rm tot}$
with respect to $A_0$ yields a relation between
the ``magnetic'' field
$B \equiv \partial_2 A_1 - \partial_1 A_2$ and the particle density
$\rho$, namely
\EQ
B= - {e\over \kappa}\,\rho
\label{2.23}
\EN
which means that a magnetic flux is attached to each particle.
Fixing the Weyl gauge $A_0=0$ and removing any residual gauge invariance
by imposing the subsidiary condition $\partial_i A^i=0$, we can solve
(\ref{2.23}) and obtain $A^i(x)$. The solution is
\EQ
A^i(x)= \sum_{I=1}^N A_I^i(\vec{x},t)\ \ ,
\label{2.24}
\EN
where
\EQ
A_I^i(\vec{x},t) \equiv A_I^i({\vec r}_1(t),...,{\vec r}_N(t))
\Big|_{{\vec r}_I={\vec x}}
\label{2.25}
\EN
with
\EQ
A^i_I({\vec r}_1,...,{\vec r}_N)
= -{e\over 2\pi \kappa}\sum_{J\not = I}
\epsilon^{ij}{{{r_j}_I-{r_j}_J}\over{\left |
{\vec r}_I-{\vec r}_J\right|^2}}
\label{2.26}
\EN
Despite the appearance, the meaning of this solution is quite
simple. In fact, the Hamiltonian corresponding to
the action $S_{\rm tot}$
is
\EQ
H'=\sum_{I=1}^N{1\over 2m}\left( {\vec p}_I - {e}~
{\vec A}_I({\vec r}_1,...,{\vec r}_N)\right)^2 +
V\left({\vec r}_1(t),...,{\vec r}_N(t)\right)\ \ .
\label{2.27}
\EN
Thus, the net effect of the Chern-Simons dynamics is to produce
a non-local vector potential ${\vec A}_I$ for each
particle. This determines a
magnetic field $B$ which vanishes
everywhere except at the particle locations,
as we can see also from (\ref{2.23}) upon using
(\ref{2.18}). Therefore the particles
effectively carry both a charge $e$ and
a magnetic flux $\phi=-e/\kappa$.
The exotic statistics is then produced by the
Aharonov-Bohm mechanism: when two particles
are exchanged they pick up a phase
because the charge of one particle moves around the flux of the other
and {\it vice versa}. To make the connection with (\ref{2.16}) more
transparent, let us observe that
\EQ
A^i_I({\vec r}_1,...,{\vec r}_N)
={e\over 2\pi\kappa}\,{\partial\over {\partial {r_i}_I}}
\sum_{J\not =I}\Theta_{IJ}\ \ ,
\label{2.28}
\EN
where $\Theta_{IJ}$ is the winding angle of particle $I$ with respect to
particle $J$ which is explicitly given by
\EQ
\Theta_{IJ} = \tan^{-1}
\left({{{r_2}_I-{r_2}_J}\over{{r_1}_I-{r_1}_J}}\right)\ \ .
\label{2.28'}
\EN
Thus the Lagrangian corresponding to
(\ref{2.27}) is
\bea
{\cal L}' &=& {\cal L} + e \,\sum _{I=1}^N
{\vec v}_I\cdot
{\vec A}_I({\vec r}_1,...,{\vec r}_N) \nonumber \\
&=&{\cal L} - {e^2\over{2\pi\kappa}}
\sum_{I<J} \left(v_I^i-v_J^i\right) {\partial\over {\partial r_I^i}}
\Theta_{IJ} \ \ ,
\label{2.29}
\ena
where we have used (\ref{2.28}) and the property
${\partial\over {\partial r_J^i}}\Theta_{IJ} = -{\partial\over
{\partial r_I^i}}
\Theta_{IJ}$. If we now observe that
\[
\left(v_I^i-v_J^i\right) {\partial\over {\partial r_I^i}}
\Theta_{IJ} ={d\over dt}\Theta_{IJ} \ \ ,
\]
we recognize that ${\cal L}'$ in (\ref{2.29})
is of the same form of ${\cal L}_F$ in (\ref{2.16})
with
\EQ
{\nu}' = {e^2\over{2\pi \kappa}}
\label{2.31}
\EN
Thus the system described by $S_{\rm tot}$ describes fermion-based
anyons with statistics ${\nu}'$ given by (\ref{2.31}).

The same results can also be obtained using the second-quantized
formalism. To see this, let us introduce
a non-relativistic fermionic matter
field $\psi({\vec x},t)$ of mass $m$ and charge $e$,
which we minimally couple
to an abelian gauge field $A_\alpha({\vec x},t)$ with a Chern-Simons
kinetic term. Thus, we consider the following action
\EQ
S = \int d^3x\left[
{\rm i}\, \psi^\dagger D_0\psi + {1\over 2m}\,\psi^\dagger
\left({D_1}^2+{D_2}^2\right)\psi + {\kappa\over 2}
\epsilon^{\alpha\beta\gamma}\,A_\alpha\,\partial_\beta \,
A_\gamma\right]\ \ ,
\label{2.32}
\EN
where $D_\alpha=\partial_\alpha +{\rm i}\,e\,A_\alpha$ is the covariant
derivative. Varying $S$ with respect to $A_\alpha$, we obtain
\EQ
\epsilon^{\alpha\beta\gamma}\,\partial_\beta A_\gamma
= {e \over \kappa}\,j^\alpha\ \ ,
\label{2.33}
\EN
where the current $j^\alpha$ is explicitly given by
\EQ
j^0 = \psi^\dagger\psi\equiv \rho
{}~~~,~~~
j^i ={{\rm i}\over{2m}}\left(\psi^\dagger D^i\psi -
\left(D^i\psi\right)^\dagger
\psi\right)\ \ .
\label{2.34}
\EN
Notice that
$\rho$ and ${\vec j}$ are respectively the second-quantized density
and current operators that correspond to (\ref{2.18}) and
(\ref{2.19}) and satisfy the
continuity equation
\EQ
\partial_0 \rho + {\vec \nabla} \cdot {\vec j} = 0 \ \ .
\label{2.35}
\EN
As is obvious from (\ref{2.33}), the Chern-Simons field strength
$(\partial_\beta A_\gamma - \partial_\gamma A_\beta)$
is completely determined
by the particle currents. Now we show that the
Chern-Simons potential itself
is not an independent degree of freedom. The $\alpha=0$ component
of (\ref{2.33}) is
\EQ
\partial_1A_2-\partial_2A_1= {e\over \kappa}\,\rho
\label{2.36}
\EN
which is simply the second-quantized
version of the quantum mechanical constraint
(\ref{2.23}). Imposing the condition
$\partial_iA^i=0$, we can solve (\ref{2.36})
and formally obtain
\EQ
A^i(x) = \epsilon^{ij}{\partial\over{\partial x^j}}
\left({e\over \kappa}\int d^2y~G({\vec x}-{\vec y})\,\rho(y)
\right)\ \ ,
\label{2.37}
\EN
where $G$ is the Green function for the Laplacian
$\Delta={\vec \nabla}\cdot{\vec\nabla}$,
satisfying
\EQ
\Delta G({\vec x}-{\vec y}) = \delta({\vec x}-{\vec y}) \ \ .
\label{2.38}
\EN
As is well known, the explicit solution to (\ref{2.38}) is
\EQ
G({\vec x}-{\vec y}) =
{1\over 2\pi} \ln\left |{\vec x}-{\vec y}\right|\ \ ,
\label{2.39}
\EN
and thus $A^i$ can be written as follows
\EQ
A^i(x) = \epsilon^{ij}{\partial\over{\partial x^j}}
\left({e\over 2\pi \kappa}\int d^2y~\ln\left |
{\vec x}-{\vec y}\right |\,\rho(y)
\right) \ \ ,
\label{2.40}
\EN
or equivalently
\EQ
A^i(x) = {e\over 2\pi \kappa}\int d^2y~{\partial\over{\partial x_i}}
\Theta({\vec x}-{\vec y})\,\rho(y)\ \ ,
\label{2.41}
\EN
where $\Theta({\vec x}-{\vec y})$ is the angle under which
${\vec x}$ is seen from ${\vec y}$, namely
\EQ
\Theta({\vec x}-{\vec y}) = \tan^{-1}\left({x_2-y_2}\over{x_1-y_1}
\right)\ \ .
\label{2.42}
\EN
In the following we will examine in detail the properties
of this angle function, but for the moment it
is enough to realize that it
is a multi-valued function so that it is necessary
to fix a cut and a reference axis in order to remove any ambiguity.
Therefore, particular care must be used in moving the derivative
$\partial/\partial x_i$ out of the integral in (\ref{2.41}) and
displaying $A^i$ as a gradient. However,
when the density $\rho$ is a sum of localized $\delta$-functions,
as is appropriate for our collection of non-relativistic point
particles, there are no
problems in moving $\partial/\partial x_i$ outside
the integral in (\ref{2.41}) and formally
write the vector potential as a pure
gauge, namely
\EQ
A^i(x)={e\over 2\pi \kappa}~{\partial\over\partial x_i}
\left(\int d^2y~
\Theta({\vec x}-{\vec y})\,\rho(y)\right )\ \ .
\label{2.43}
\EN
Because of translational invariance of the angle function (\ref{2.42}),
it is harmless to
shift the density $\rho(y)$ by a constant $\rho_0$ and write
in general
\EQ
A^i(x)={\partial\over\partial x_i}\Lambda(x)\ \ ,
\label{2.44}
\EN
where
\EQ
\Lambda(x) = {e\over 2\pi \kappa}~\int d^2y~
\Theta({\vec x}-{\vec y})\,\left(\rho(y)-\rho_0\right)\ \ .
\label{2.45}
\EN
Now we show that also $A_0$ is pure gauge.
To this aim, let us first observe
that the space-components of
(\ref{2.33}) are
\EQ
\partial_iA_0-\partial_0A_i= -{e\over \kappa}
\epsilon_{ij}j^j\ \ .
\label{2.46}
\EN
Then, upon acting with $\partial/\partial x_i$
and recalling that $\partial_iA^i=0$,
we easily get, after an integration by parts,
\EQ
A_0(x) = -{e\over \kappa}
\int d^2y~\epsilon^{ij}{\partial\over \partial x^j}
G({\vec x}-{\vec y}) ~j_i(y)\ \ ,
\label{2.47}
\EN
or equivalently
\EQ
A_0(x) = -{e\over 2\pi \kappa}
\int d^2y~{\partial\over\partial x_i}
\Theta({\vec x}-{\vec y})~j_i(y)\ \ .
\label{2.47'}
\EN
If we observe that ${\partial\over\partial x^i}\Theta({\vec x}-{\vec y})
=-{\partial\over\partial y^i}\Theta({\vec x}-{\vec y})$, and
after a further integration
by parts we use the continuity equation for the current, we finally get
\EQ
A_0(x)={e\over 2\pi \kappa}
\int d^2y ~\Theta({\vec x}-{\vec y})~\partial_0
\left(\rho(y)-\rho_0\right) = \partial_0\Lambda(x)\ \ .
\label{2.48}
\EN
Eqs. (\ref{2.44}) and (\ref{2.48}) can be combined
into a single covariant expression
$A_\alpha(x)=\partial_\alpha \Lambda(x)$ which exhibits the fact that
the Chern-Simons potential in this model is pure gauge, {\it albeit}
of a non-standard form because of the multivaluedness
of the gauge function $\Lambda(x)$. This feature implies that
$\partial_\alpha \Lambda(x)$ is
not a trivial quantity: indeed, the vector potential
(\ref{2.44}) describes point-like
``magnetic'' fluxes localized on each particle, as one can see
by computing the gauge invariant quantity
\[
\oint_{{\cal C}_I} d{\vec \ell}\cdot {\vec A}\ \ ,
\]
where ${\cal C}_I$ is a closed path encircling particle $I$.

Let us now quantize the action (\ref{2.32}) by imposing equal-time
anticommutation relations on the fermionic field $\psi$, namely
\EQ
\Big\{\psi({\vec x},t)\,,\,\psi^\dagger({\vec y},t)\Big\} =
\delta({\vec x}-{\vec y})~~~,~~~
\Big\{\psi({\vec x},t)\,,\,\psi({\vec y},t)\Big\}
=\Big\{\psi^\dagger({\vec x},t)\,,\,\psi^\dagger({\vec y},t)\Big\}=0\ \ .
\label{2.49}
\EN
Since the Chern-Simon field is a
functional of the density $\rho=\psi^\dagger\,\psi$,
the commutation relation relations between $A_\alpha$ and $\psi$ are not
trivial, but they can be easily worked out using (\ref{2.49}),
(\ref{2.44}) and (\ref{2.48}). Once this is done, it is
possible to obtain the first-quantized Hamiltonian that is equivalent
to the second-quantized action (\ref{2.32}). As a result, one finds
precisely the Hamiltonian (\ref{2.27}) with $V=0$, so that one can say
that the system described
by (\ref{2.32}) is a system of fermion-based anyons
of statistics ${\nu}'$ as in (\ref{2.31})
(for simplicity hereinafter ${\nu}'$
will be denoted simply by $\nu$).
We refer the reader
to \cite{L} for the explicit derivation of this result. Here instead we
want to point out that there is another way of recognizing
that our system (\ref{2.32})
is actually a system of anyons.
In fact, since the Chern-Simons field
is formally
pure gauge, it can be removed with a (singular) gauge transformation of
parameter $\Lambda$, namely
\EQ
A_\alpha \longrightarrow A'_\alpha
= A_\alpha - \partial_\alpha \Lambda =0 \ \ .
\label{2.50}
\EN
Under this transformation
the matter field $\psi$ transforms according to
\bea
\psi(x)\longrightarrow\psi'(x) &=&
{\rm e}^{{\rm i}\,e\,\Lambda(x)}~\psi(x)
\nonumber \\
&=& {\rm e}^{{\rm i}\,\nu\int d^2y~
\Theta({\vec x}-{\vec y})\,\big(\psi^\dagger(y)
\psi(y)-
\rho_0\big)}~\psi(x)\ \ ,
\label{2.51}
\ena
while the covariant derivatives $D_\alpha$ turn into
ordinary derivatives $\partial_\alpha$.
Thus the action (\ref{2.32}) becomes simply
\EQ
S'=\int d^3x\left[
{\rm i} {\psi '}^\dagger
\partial_0{\psi '} + {1\over 2m}{\psi '}^\dagger
\,\Delta\,{\psi '}\right] \ \ .
\label{2.52}
\EN
This is a free action. However, it should be realized that
the non-trivial effects of the Chern-Simons
dynamics are actually hidden in the new matter fields
${\psi '}$ which do
not satisfy any more canonical anticommutation relations like
(\ref{2.49}).
In fact, as we can see with a few simple algebraic manipulations
using (\ref{2.51}) and (\ref{2.49}), we have
\EQ
{\psi '}({\vec x},t)~{\psi '}({\vec y},t)=-
{\rm e}^{\,-{\rm i}\,\nu\,\big[\Theta({\vec x}-{\vec y})-
\Theta({\vec y}-{\vec x})\big]}~{\psi '}({\vec y},t)~
{\psi '}({\vec x},t)\ \ .
\label{2.53}
\EN
This is an example of a braiding relation, because the phase factor in
the right hand side
is different for different relative positions of ${\vec x}$ and
${\vec y}$.
To make this statement more precise, let us recall that the
multi-valued angle function $\Theta({\vec x}-{\vec y})$ is
unambiguously defined only after one puts a cut in the
plane and fixes a reference
axis. For example \cite{LS} one can choose
as a cut the negative $x$-axis and
measure the angles
starting from the positive $x$-axis. With this choice
all angles are in the interval $~[-\pi,\pi)~$.
Then it is not difficult to find that
\EQ
\Theta({\vec x}-{\vec y}) - \Theta({\vec y}-{\vec x}) =
\left\{\begin{array}{cc}
 \pi \,{\rm sgn}(x_2-y_2)  &~~~~~{\rm for}~~~x_2\not=y_2\ \ ,\\
 \pi \,{\rm sgn}(x_1-y_1)  &~~~~~{\rm for}~~~x_2=y_2 \ \ .
\\
\end{array}\right.
\label{2.54}
\EN
Using this result in (\ref{2.53}), it
becomes clear that ${\psi '}$ behaves as
an anyonic field. Furthermore, since ${\psi '}$ depends also on
the cut of the angle function, it is a {\it non local} operator.
It is precisely
for this non local character that the
exchange relations
(\ref{2.53}) are unambiguous. In fact,
if $x_2>y_2$ (or if $x_2=y_2$ and
$x_1>y_1$), there is only one way in
which ${\psi '}({\vec y},t)$ can move
around ${\psi '}({\vec x},t)$ without
crossing its cut, namely counterclockwise
(see Fig. 1a). In this case, $\Theta({\vec x}-{\vec y}) -
\Theta({\vec y}-{\vec x}) =
\pi$ and (\ref{2.53}) becomes
\EQ
{\psi '}({\vec x},t)~{\psi '}({\vec y},t)=-q^{-1}
{}~{\psi '}({\vec y},t)~{\psi '}({\vec x},t)\ \ ,
\label{2.55}
\EN
where
\EQ
q={\rm e}^{\,{\rm i}\,\nu\,\pi}
\label{2.56}
\EN
On the contrary, if $x_2<y_2$ (or if $x_2=y_2$ and
$x_1<y_1$) the field ${\psi '}({\vec y},t)$ must move
around ${\psi '}({\vec x},t)$ clockwise in
order not to cross its cut
(see Fig. 1b). In this case $\Theta({\vec x}-{\vec y})
- \Theta({\vec y}-{\vec x}) =
-\pi$, and (\ref{2.53}) becomes
\EQ
{\psi '}({\vec x},t)~{\psi '}({\vec y},t)=-q
{}~{\psi '}({\vec y},t)~{\psi '}({\vec x},t)\ \ .
\label{2.57}
\EN
Thus, fixing the cut for the angle function
automatically
fixes also the orientation of the exchange of
two fields ${\psi '}$ and
determines unambiguously their braiding properties.
In conclusion, we can say that the
gauge transformation (\ref{2.51}) transmutes fermions
into anyons of statistics $\nu$
and so it can be considered as the two-dimensional generalization
of the Jordan-Wigner transformation \cite{JW}, which in one dimension
transmutes fermions into bosons.

Before extending further our analysis, a
few remarks are in order. First of all, the fact that
the transformation
from $\psi$ to ${\psi '}$ in (\ref{2.51}) involves expressly
the angle function $\Theta ({\vec x}-{\vec y})$ is a consequence
of fixing the transverse gauge $\partial_iA^i=0$ on the Chern-Simons
field. Different gauge fixings would lead to continuous
deformations of the angle function (still denoted by
$\Theta ({\vec x}-{\vec y})$)
which maintain
the same multivalued properties of angles. However, in computing the
exchange relations between two transformed fields (see for instance
(\ref{2.53}))
only one particular combination
of these functions always appears, namely
$\Theta ({\vec x}-{\vec y}) - \Theta ({\vec y}-{\vec x})$.
Thus, since the particles created by ${\psi '}$
in ${\vec x}$ and ${\vec y}$ are indistinguishable, one
is led to choose only those
gauges for which
\[
\Theta ({\vec x}-{\vec y}) - \Theta ({\vec y}-{\vec x}) = \pm \pi
\]
in such a way that the braiding relations are like (\ref{2.55}) or
(\ref{2.57}), {\it i.e.}
with only constant phase factors involved. As we
have seen, the transverse gauge is one of these gauges (perhaps
the most obvious), but of course other choices are possible.
We will exploit this freedom in the following sections.

We conclude this review of anyons by recalling
that the Chern-Simons construction of fractional statistics and the
Jordan-Wigner transformation transmuting fermions into
anyons can be realized also if the two-dimensional space is
discrete, {\it i.e.} on a lattice $\Omega$.  For traditional
reasons we denote the fermions on the lattice by
$c({\vec x})$ and $c^\dagger({\vec x})$ and
for simplicity we do not write
the time dependence.
They satisfy the
following standard equal-time anticommutation relations
\EQ
\Big\{c({\vec x})\,,\,c^\dagger({\vec y})\Big\} =
\delta_{{\vec x},{\vec y}}~~~,~~~
\Big\{c({\vec x})\,,\,c({\vec y})\Big\}
=\Big\{c^\dagger({\vec x})\,,\,c^\dagger({\vec y})\Big\}=0\ \ ,
\label{2.60}
\EN
where $\delta_{{\vec x},{\vec y}}$ is the lattice $\delta$-function.

If we want to generalize the Jordan-Wigner transformation to the lattice,
we must first define the angle function $\Theta ({\vec x},{\vec y})$
for ${\vec x}$ and ${\vec y} \in \Omega$.
The definition of the
lattice angle function requires some care \cite{LMES,LS},
but it is not
difficult and can be regarded as a straightforward generalization
of the continuum angle function considered so far.
The only point that we want to mention here is that the cut that must
be fixed to remove any ambiguity in the definition,
has to be chosen on a suitably defined dual lattice $\Omega^*$.
(We refer the reader to \cite{LS} for a
thorough discussion of this issue.)
Hence, also the reference point from which the angles are measured
is always a point of $\Omega^*$.
For example we can choose as cuts the lines $\gamma$'s represented in
Fig. 2 for a few points of $\Omega$. With this choice, the angles
are in the interval $~[-\pi,\pi)$ and the corresponding
lattice function
$\Theta ({\vec x},{\vec y})$ satisfies
\EQ
\Theta({\vec x},{\vec y}) - \Theta({\vec y},{\vec x}) =
\left\{\begin{array}{cc}
 \pi \,{\rm sgn}(x_2-y_2)  &~~~~~{\rm for}~~~x_2\not=y_2\ \ ,\\
 \pi \,{\rm sgn}(x_1-y_1)  &~~~~~{\rm for}~~~x_2=y_2 \ \ .
\\
\end{array}\right.
\label{2.54'}
\EN
This relation is very important because it allows to
establish an ordering relation on the lattice.
In fact, given two distinct points ${\vec x}$
and ${\vec y}\in \Omega$, we
can posit
\bea
{\vec x}>{\vec y} & \Longleftrightarrow &
\Theta({\vec x},{\vec y}) -
\Theta({\vec y},{\vec x}) =
\pi \ \ ,
\nonumber \\
{\vec x}<{\vec y} & \Longleftrightarrow &
\Theta({\vec x},{\vec y}) -
\Theta({\vec y},{\vec x}) =
-\pi \ \ .
\label{2.a}
\ena
This definition is
unambiguous and endows $\Omega$ with an ordering
relation enjoying all the correct properties.
This ordering relation will play a crucial
role in the following sections when we will
discuss the connection between anyons and quantum groups.

After the lattice angle function is defined, we can define
the so-called disorder operators \cite{F} according to
\EQ
K({\vec x}) = {\rm e}^{\,{\rm i}\,
\nu\sum\limits_{{\vec y}\not = {\vec x}}
\Theta({\vec x},{\vec y})\,\left(
c^\dagger({\vec y})c({\vec y})-\rho_0\right)} \ \ ,
\label{2.61}
\EN
where $\rho_0$ is a constant to be fixed later,
and prove that
\bea
K({\vec x})~c({\vec y}) &=&
{\rm e}^{\,-{\rm i}\,\nu\,\Theta({\vec x},{\vec y})}
{}~c({\vec y})~K({\vec x}) \ \ , \nonumber \\
K({\vec x})~c^\dagger({\vec y}) &=&
{\rm e}^{\,{\rm i}\,\nu\,
\Theta({\vec x},{\vec y})}~c^\dagger({\vec y})~K({\vec x})\ \ , \
\nonumber \\
K({\vec x})~K({\vec y}) &=&
K({\vec y})~K({\vec x})
\label{2.62}
\ena
for all ${\vec x}$ and ${\vec y}\in \Omega$.
Then, using the disorder operators $K({\vec x})$,
we define the lattice anyons
as follows
\EQ
a({\vec x}) = K({\vec x})~c({\vec x}) =
{\rm e}^{\,{\rm i}\,
\nu\sum\limits_{{\vec y}\not = {\vec x}}
\Theta({\vec x},{\vec y})\,\left(
c^\dagger({\vec y})c({\vec y})-\rho_0\right)}~c({\vec x})\ \ .
\label{2.63}
\EN
This is the generalized Jordan-Wigner
transformation and is nothing but the lattice version of the gauge
transformation (\ref{2.51}). It is rather easy to show that the
anyonic operators $a$ satisfy the following braiding relations
\bea
a({\vec x})~a({\vec y})&+&q^{-1}~
a({\vec y})~a({\vec x})=0
\ \ ,\nonumber \\
a({\vec x})~a^\dagger({\vec y})
&+&q~a^\dagger({\vec y})~a({\vec x})=0
\ \ , \nonumber \\
a^\dagger({\vec x})~
a^\dagger({\vec y})&+&q^{-1}~
a^\dagger({\vec y})~
a^\dagger({\vec x})=0 \ \ , \nonumber \\
a^\dagger({\vec x})~a({\vec y})
&+&q~a({\vec y})~a^\dagger({\vec x})=0
\label{2.64}
\ena
for all ${\vec x} > {\vec y}$. We notice that the last two relations are
simply the hermitian conjugate of the first two, since $q^*=q^{-1}$.
For completeness we also point out that
\EQ
a({\vec x})^2=
a^\dagger({\vec x})^2=0\ \ ,
\label{2.65}
\EN
and
\EQ
a({\vec x})~a^\dagger({\vec x})
+a^\dagger({\vec x})~a({\vec x})=1 \ \ .
\label{2.66}
\EN
Thus, the anyonic operators $a$ obey the Pauli exclusion
principle, and at the same point
satisfy standard anticommutation relations without any phase factor,
like their parent fermions.

We end our discussion by recalling that different choices of
the cut and of the reference axis
for the angle function lead to different kinds of disorder
operators, and hence to different kinds of anyons. For example,
instead of the cuts $\gamma$ considered so far, we can choose the
cuts $\delta$ along the positive $x$-axis and
measure the angles starting from the negative $x$-axis
(see Fig. 3). Comparing Figs. 2 and 3, it is clear that
$\delta$ and $\gamma$ are related to each other by a sort
of parity transformation. We denote by
${\tilde \Theta}({\vec x},{\vec y})$
the new angle function which is equal to $\Theta({\vec y},{\vec x})$,
and therefore satisfies
\EQ
{\tilde\Theta}({\vec x},{\vec y}) - {\tilde\Theta}({\vec y},{\vec x}) =
\left\{\begin{array}{cc}
 -\pi  &~~~~~{\rm for}~~~{\vec x}>{\vec y}\ \ ,\\
  \pi   &~~~~~{\rm for}~~~{\vec x}<{\vec y}\ \ ,
\\
\end{array}\right.
\label{2.66'}
\EN
and by ${\tilde a}$ the anyonic operators
that can be constructed thereupon, {\it i.e.}
\EQ
{\tilde a}({\vec x}) = {\rm e}^{\,{\rm i}\,
\nu\sum\limits_{{\vec y}\not = {\vec x}}
{\tilde \Theta}({\vec x},{\vec y})\,\left(
c^\dagger({\vec y})c({\vec y})-\rho_0\right)}~c({\vec x})\ \ .
\label{2.67}
\EN
The operators ${\tilde a}$
are again fermion-based anyons of statistics $\nu$, and
satisfy the braiding relations (\ref{2.64}) with $q$
replaced by $q^{-1}=q^*$.
Indeed, ${\tilde a}$ is the parity transformed of $a$, and thus
braids in the opposite way (see Fig. 4 in comparison
with Fig. 1). We will use both anyons of
type $a$ and anyons of type
${\tilde a}$ in the subsequent sections.

Finally, we observe that the anyonic operators defined in
(\ref{2.63}) and (\ref{2.67}) have nothing to do with the
so-called $q$-oscillators \cite{qO}, despite some formal analogies.
This is so for several reasons. First of all,
the $q$-oscillators can be defined in any dimensions and are not related
to the braid group, whereas anyons are strictly two-dimensional objects.
Secondly, the $q$-oscillators are local operators and are
characterized by $q$-commutation relations quite different from
(\ref{2.66}), whilst anyons are
intrinsically non-local due to their braiding properties.
This non-locality, essential to distinguish whether anyons
are exchanged clockwise or anticlockwise, allows to define a
natural ordering among the particles, which in turn will be essential
for the realization of the quantum
groups presented in Sections 4 and 5.

\vspace{1cm}

%
%

\sect{Deformed Lie Algebras}
\vspace {0.7cm}

As we mentioned in the introduction, both anyons and deformed
Lie algebras are deeply related to the braid group,
and so it is natural to conjecture
the existence of a direct relation between them. In
this section and in the following ones we prove that
this relation does indeed exist, and in particular we
show that representations of deformed Lie algebras can be constructed
using the non-local anyonic operators $a$ and
$\tilde a$ previously defined.
In order to illustrate this fact, we first recall some
fundamental properties of deformed Lie algebras and
briefly discuss their relation with the ordinary ({\it i.e.} undeformed)
ones.

By construction, a deformed Lie algebra is a deformation of an ordinary
Lie algebra to which it reduces when
the deformation parameter, usually denoted by $q$, goes to $1$.
Given an ordinary Lie algebra $G$ of rank $r$, its $q$-deformation
${\cal U}_q(G)$ is characterized by
generalized commutation relations which, in the
Chevalley basis, are
\subequations
\bea
\Big[H_i~,~H_j\Big]&=&0 \ \ ,
\label{3.1a} \\
\Big[H_i~,~E_j^{\pm}\Big]&=&\pm a_{ij}~E_j^{\pm} \ \ ,
\label{3.1b} \\
\Big[E_i^+~,~E_j^-\Big]&=&\delta_{ij}~\left[H_i \right]_{q_i}\ \ ,
\label{3.1c} \\
\sum_{\ell=0}^{1-a_{ij}}(-1)^\ell{{1-a_{ij}}\brack \ell}_{q_i} &&
\!\!\!\!\!\!\!\!\!\!\left(E_{i}^{\pm}\right)^{1-a_{ij}-\ell}~E_{j}^{\pm}
{}~\left(E_{i}^{\pm}\right)^\ell=0 \ \ ,
\label{3.1d}
\ena
\endsubequations
where $H_i$ ($i=1,...,r$) are the generators of the Cartan subalgebra
of $G$,
$E_i^{\pm}$
are the step operators corresponding to the simple root $\alpha_i$,
and $a_{ij}$ are the elements the Cartan matrix, {\it i.e.}
\[
a_{ij}=\left\langle \alpha_i,\alpha_j \right\rangle =
2~{{(\alpha_i,\alpha_j)}\over{(\alpha_i,\alpha_i)}}
\]
(see Tab. 1).
In eqs. (3.1) we have used the standard notations
\bea
{[x]}_q&=&{\frac {q^x-q^{-x}}{q-q^{-1}}}\ \ ,
\nonumber \\
{{m\brack n}}_q&=&{\frac {{[m]}_q!}{{[m-n]}_q!~{[n]}_q!}} \ \ ,
\label{3.3} \\
{[m]}_q!&=&{[m]}_q\,{[m-1]}_q\cdots{[1]}_q\ \ .
\nonumber
\ena
Furthermore, $q_i\equiv q^{{1\over 2}\,(\alpha_i,\alpha_i)}$
so that
\[
{q_i}^{a_{ij}}~=~{q_j}^{a_{ji}}\ \ .
\]

To complete the definition of ${\cal U}_q(G)$, we recall that
the comultiplication $\Delta$,
the antipode $S$ and the co-unit $\epsilon$ are given by
\subequations
\bea
\Delta(H_i) &=&
 H_i \otimes{\bf 1}+{\bf 1}\otimes H_i \ \ , \label{3.6a} \\
\Delta(E_i^{\pm})&=&
E_i^{\pm}\otimes {q_i}^{{H_i}/2}+
{q{}_i}^{-{H_i}/2}\otimes E_i^{\pm}\ \ , \label{3.6b} \\
S({\bf 1}) &=&
{\bf 1}~~,~~~~S(H_i)=-H_i \ \ , \label{3.6c} \\
S(E_i^{\pm })&=&
-{q_i}^{{H_i}/2}~ E_i^{\pm}~{q_i}^{-{H_i}/2} \ \ ,
\label{3.6d} \\
\epsilon({\bf 1}) &=&
 1 ~~,~~~~ \epsilon(H_i) =\epsilon(E_i^{\pm})=0  \ \ .
\label{3.6e}
\ena
\endsubequations
Notice that for complex $q$ consistency of eqs. (\ref{3.6b}) and (\ref{3.6d})
implies
that $E_i^-$ is not
the adjoint of $E_i^+$, but
\EQ
\left. E_i^-\right|_q = \left(\left. E_i^+\right|_{q^*}\right)^\dagger
\ \ .
\label{3.6'}
\EN
It is easy to
realize that when $q\to 1$, eqs. (3.1), (3.3) and (\ref{3.6'})
reduce to those appropriate for an ordinary Lie algebra, thus clearly
exhibiting the fact that ${\cal U}_q(G)$ is a deformation of $G$.

When $G$ is a classical Lie algebra ({\it i.e.} belonging to the $A$,
$B$, $C$ and $D$ series), the connection between deformed
and undeformed Lie algebras is even closer \cite{FMS}. In fact
there exists a set of non trivial representations of ${\cal U}_q(G)$
that do not depend on $q$ and are therefore common both to the deformed
and undeformed enveloping algebras
\footnote{Actually this property
holds also for $E_6$ and $E_7$, but not for the remaining exceptional
algebras. The whole discussion of this section can thus be
referred also to ${\cal U}_q(E_6)$ and  ${\cal U}_q(E_7)$.}.
This happens when all the $SU(2)$ subalgebras relative to the simple
roots of $G$ are in the spin-$0$ or spin-$1/2$ representation.
We call ${\Re}_{(0,1/2)}$
the set of all representations with this property.

Let us denote by $h_i$ and $e_i^{\pm}$ the generators
$H_i$ and $E_i^\pm$ in a representation
belonging to ${\Re}_{(0,1/2)}$; then
the following two properties hold
\begin{enumerate}
\item{} the eigenvalues of $h_i$ ({\it i.e.} the Dynkin
labels of any weight) are either $0$ or $\pm 1$, and
\item{} $(e_i^{\pm})^2~=~0$.
\end{enumerate}

Property $1$ implies that
\[
\left[h_i \right]_{q_i}~=~ h_i
\]
for any value of $q$. On the other hand,
because of property $2$,
the deformed Serre relation (\ref{3.1d}) is
identically satisfied for all $i$ and $j$ such that $a_{ij}=-2$,  and
simply becomes
\[
-(q_i~+~q_i^{-1})~e_i^{\pm}~e_j^{\pm}~e_i^{\pm}=0
\]
for all $i$ and $j$ such that $a_{ij}=-1$.
Finally, if $a_{ij}=0$ eq. (\ref{3.1d}) reduces to
\[
\Big[e_i^{\pm}~,~e_j^{\pm}\Big]=0
\]
for all values of $q$.
\par
These facts show that for the representations
belonging to ${\Re}_{(0,1/2)}$
the deformed commutation relations (3.1)
are actually independent of the deformation parameter $q$ and
therefore coincide with the undeformed ones.

The set ${\Re}_{(0,1/2)}$ is not empty. In fact,
for any classical Lie algebra $G$ the
fundamental representations (see Fig. 5) certainly belong to
${\Re}_{(0,1/2)}$ (by fundamental representation we mean an
irreducible representation such that any other representation
can be constructed from it by taking
tensor products, or, equivalently, by repeated use
of comultiplication). Thus, the fundamental representation of
$G$ can be interpreted also as a representation of ${\cal U}_q(G)$.

Let us now introduce
an {\it ordered} set $\Omega$ whose elements we denote
by ${\vec x}$. Later on we will identify $\Omega$
with a two-dimensional lattice,
but for the moment this interpretation is not needed.
To each point ${\vec x} \in \Omega$ we
assign a fundamental representation of $G$ and denote by $h_i({\vec x})$
and $e_i^{\pm}({\vec x})$ the corresponding generators.
The previous discussion
assures that these local generators
satisfy the following generalized
commutations relations
\subequations
\bea
\Big[h_i({\vec x})~,~h_j({\vec y})\Big]&=&0 \ \ ,
\label{3.10a} \\
\Big[h_i({\vec x})~,~e_j^{\pm}({\vec y})\Big]&=&
\pm \delta({\vec x},{\vec y})~ a_{ij}~e_j^{\pm}({\vec x}) \ \ ,
\label{3.10b} \\
\Big[e_i^{+}({\vec x})~,~e_j^{-}({\vec y})\Big]&=&
\delta({\vec x},{\vec y})~\delta_{ij}~
\big[h_i({\vec x})\big]_{q^{}_i} \ \ ,
\label{3.10c} \\
\sum_{\ell=0}^{1-a_{ij}}(-1)^\ell{{1-a_{ij}}
\brack \ell}_{q^{}_i}  &&
\!\!\!\!\!\!\!\!\!\!\left(e_i^{\pm}({\vec x})\right)^{1-a_{ij}-\ell}
{}~e_j^{\pm}({\vec x})
{}~\left(e_i^{\pm}({\vec x})\right)^\ell=0 \ \ ,
\label{3.10d} \\
\Big[e_i^{\pm}({\vec x})~,~e_j^{\pm}({\vec y})\Big]&=&
0 ~~~~{\rm for}~~~{\vec x}\not={\vec y}\ \ .
\label{3.10e}
\ena
\endsubequations
It should be clear that the relations (3.5) are just
formally deformed and are actually independent of $q$ because
$h_i({\vec x})$ and $e_i^{\pm}({\vec x})$ are in
a representation belonging
to ${\Re}_{(0,1/2)}$; however, for our
later discussion, it is useful to write them as
deformed commutation relations.

Now we can make an iterated use of the coproduct $\Delta$ of
${\cal U}_q(G)$ as given in (\ref{3.6a}) and
(\ref{3.6b})\footnote{Notice that for $q\neq 1$ the coproduct is not
cocommutative and for this reason we require that $\Omega$ is ordered.},
and combine all the local
representations to yield the global generators
\EQ
H_i =\sum_{{\vec x}\in \Omega} H_i({\vec x})
{}~~~~,~~~~
E_i^{\pm} =
\sum_{{\vec x}\in \Omega}E_i^{\pm}({\vec x}) \ \ ,
\label{3.11}
\EN
where
\subequations
\bea
H_i({\vec x}) &=& h_i({\vec x})\ \ ,
\label{3.12a} \\
E_i^{\pm}({\vec x}) &=&\prod_{{\vec y}<{\vec x}}
{q^{}_i}^{-h_i({\vec y})/2} ~
e_i^{\pm}({\vec x})~
\prod_{{\vec z}>{\vec x}}{q^{}_i}^{h_i({\vec z})/2}\ \ .
\label{3.12b}
\ena
\endsubequations
The consistency between product and coproduct implies
that the generators $H_i$ and $E_i^{\pm}$ in (\ref{3.11})
are a ``higher spin'' representation of ${\cal U}_q(G)$ and thus
satisfy eqs. (3.1), as one can see also with a direct check
Moreover, from (\ref{3.10b}) and (\ref{3.12b}) it is not difficult to
prove that
\EQ
E_i^{\pm}({\vec x})~E_j^{\pm}({\vec y})=\left\{
\begin{array}{cc}
{q_i}^{\pm a_{ij}}~~ E_j^{\pm}({\vec y})~E_i^{\pm}({\vec x})
&~~~~~{\rm for}~~~{\vec x}>{\vec y}\ \ ,\\
{q_i}^{\mp a_{ij}}~~ E_j^{\pm}({\vec y})~E_i^{\pm}({\vec x})
&~~~~~{\rm for}~~~{\vec x}<{\vec y} \ \ .
\\
\end{array}\right.
\label{3.14}
\EN
If ${\vec x}$ and ${\vec y}$ are points of a two-dimensional
lattice $\Omega$, then eqs. (\ref{3.14}) can be interpreted as
(generalized) braiding relations similar to those satisfied
by the anyonic operators defined in Section 2.
Moreover, as we have already observed, the set $\Omega$ has to be ordered
and anyons can naturally provide an ordering relation.
Therefore, it is not unconceivable
to conjecture a close connection between anyons and quantum groups.
In the next two sections we prove the existence of such a connection
by constructing explicitly the generators $H_i({\vec x})$ and
$E_i^\pm({\vec x})$ in terms of the anyonic operators $a({\vec x})$ and
${\tilde a}({\vec x})$.

\vspace {1cm}

%
%

\sect{Anyonic Construction of Deformed Classical Lie Algebras}
\vspace{0.7cm}

It has been known for a long time
that any classical Lie algebra $G$ can be
constructed \`a la Schwinger in a rather straightforward way
using fermionic (or bosonic)
oscillators. More recently \cite{LS,FMS,CM} we
have found that
replacing fermions with anyons of statistics
$\nu$
in the Schwinger construction of $G$, one gets a realization
of the deformed algebra ${\cal U}_q(G)$ with
$q= e^{{\rm i}\,\pi\,\nu}$.
In this section we present this result for the algebras of the
$A$, $B$ and $D$ series, whereas we postpone to the next
section the discussion of the algebras of the $C$ series,
since these are slightly more complicated.

In the fermionic Schwinger construction of $G$ one makes use of many
copies of a set of independent fermionic oscillators
satisfying standard anticommutation relations. We label
each copy by a vector ${\vec x}$ and assume
that the set of all
these vectors defines a two-dimensional lattice $\Omega$.
For each point ${\vec x}\in\Omega$ we construct the local
generators $h_i({\vec x})$ and $e^\pm_i({\vec x})$ of $G$, and then
we sum these over the whole lattice to get the global ones, namely
\EQ
H_i =\sum_{{\vec x}\in \Omega} h_i({\vec x})
{}~~~~,~~~~
E_i^{\pm} =
\sum_{{\vec x}\in \Omega}e_i^{\pm}({\vec x}) \ \ .
\label{4.0}
\EN
Notice that these sums correspond simply to a repeated use of the
comultiplication of $G$.

Let us now recall some details of this construction. For the algebra
$A_r$ one introduces in each point ${\vec x}$ a set of
$(r+1)$ independent fermionic oscillators $c_i({\vec x})$,
and then defines the local generators
\subequations
\bea
h_i({\vec x})&=&n_i({\vec x})
-n_{i+1}({\vec x})\ \ ,
\label{4.1a} \\
e_i^+({\vec x})&=&c_i^\dagger({\vec x})
\,c_{i+1}({\vec x})\ \ ,
\label{4.1b} \\
e_i^-({\vec x})&=&c_{i+1}^\dagger({\vec x})
\,c_i({\vec x})\ \ ,
\label{4.1c}
\ena
\endsubequations
where $i=1,...,r\,$, and $n_i({\vec x})
=c_i^\dagger({\vec x})c_i({\vec x})$
is the standard number operator.

For the algebras $B_r$
and $D_r$, one introduces instead $r$ independent fermionic oscillators
for each ${\vec x}$. In particular for the algebra $B_r$,
$h_i({\vec x})$ and $e^\pm_i({\vec x})$ for $i=1,...,r-1$ are as
in (4.2), while $h_r({\vec x})$ and $e^\pm_r({\vec x})$
are defined as
\subequations
\bea
h_r({\vec x})&=&2n_r({\vec x})-1\ \ ,
\label{4.2a} \\
e_r^+({\vec x})&=&c_r^\dagger({\vec x})
\,{\cal S}({\vec x})\ \ ,
\label{4.2b} \\
e_r^-({\vec x})&=&c_r({\vec x})
\,{\cal S}({\vec x})\ \ ,
\label{4.2c}
\ena
\endsubequations
where
\[
{\cal S}({\vec x})=\prod_{{\vec y}<{\vec x}}\prod_{i=1}^r
(-1)^{n_i({\vec y})}
\ \
\]
is a sign factor that must be introduced to make the generators
commute at different points.

For the algebra $D_r$, $h_i({\vec x})$ and $e^\pm_i({\vec x})$
for $i=1,...,r-1$
are again given by (4.2), whereas
$h_r({\vec x})$ and $e^\pm_r({\vec x})$ are as follows
\subequations
\bea
h_r({\vec x})&=&n_{r-1}({\vec x})+
n_r({\vec x})-1 \ \ ,
\label{4.3a} \\
e_r^+({\vec x})&=&c_r^\dagger({\vec x})
\,c_{r-1}^\dagger({\vec x})\ \ ,
\label{4.3b} \\
e_r^-({\vec x})&=&c_{r-1}({\vec x})
\,c_r({\vec x})\ \ .
\label{4.3c}
\ena
\endsubequations

It is a very easy task to check that the
generators $h_i({\vec x})$ and $e_i^\pm({\vec x})$ defined
in this way satisfy
the commutation relations (3.5) with the
appropriate Cartan matrices and $q=1$.
Moreover, one can easily realize that properties $1$ and $2$ discussed
in Section 3 hold. In fact, due to the fermionic character
of the oscillators $c({\vec x})$, the eigenvalues of the
Cartan generators $h_i({\vec x})$
can only be either 0 or $\pm1$, and all step operators
$e_i^\pm({\vec x})$
have a vanishing square. Thus, the representations of
$A_r$, $B_r$ and $D_r$ constructed in this way belong to the
set ${\Re}_{(0,1/2)}$
and are fundamental ones. As mentioned before, all other
representations of these algebras can be obtained
by a repeated use
of the coproduct, that is by summing the local generators
over all points
of $\Omega$ (see (\ref{4.0})).

To obtain the
representations of the deformed Lie algebras, one could follow
the standard procedure and make use of the comultiplication
$\Delta$ defined in (\ref{3.6a}) and (\ref{3.6b}) and combine the local
representations as in (3.7). However, it is possible
to obtain a representation of the deformed Lie algebras also in an
alternative way, namely by replacing fermions with anyons in the
Schwinger construction. Because of eq. (\ref{3.6'}), both the
oscillators $a({\vec x})$ and
${\tilde a}({\vec x})$ defined in (\ref{2.63}) and (\ref{2.67})
must be used.
More precisely, let us define the anyonic generator densities
$H_i^+({\vec x})$ and  $E_i^{\pm}({\vec x})$ as
\subequations
\bea
H_i({\vec x})&=&N_i({\vec x}) - N_{i+1}({\vec x}) \ \
\label{4.6a} \ \ , \\
E_i^+({\vec x})&=&a_i^\dagger({\vec x})\,
a_{i+1}({\vec x}) \ \ ,
\label{4.6b} \\
E_i^-({\vec x})&=&{\tilde a}_{i+1}^\dagger({\vec x})\,
{\tilde a}_i({\vec x}) \ \ ,
\label{4.6c}
\ena
\endsubequations
where
\[
N_i({\vec x})=a_i^\dagger({\vec x})a_{i}({\vec x})=
{\tilde a}_i^\dagger({\vec x}){\tilde a}_{i}({\vec x})=
n_i({\vec x})
\]
$i=1,...,r$ for $A_r$, and  $i=1,...,r-1$ for $B_r$ and $D_r$.
The remaining generators
$E_r^{\pm}({\vec x})$
and $H_r({\vec x})$ for $B_r$ and $D_r$ are defined
respectively as
\subequations
\bea
H_r({\vec x})&=&2\,N_r({\vec x}) - 1 \ \ ,
\label{4.7a} \\
E_r^+({\vec x}) &=&a_{r}^\dagger({\vec x})~
{\cal S}({\vec x}) \ \ ,
\label{4.7b} \\
E_r^-({\vec x}) &=&{\tilde a}_{r}({\vec x})~
{\cal S}({\vec x}) \ \ ,
\label{4.7c}
\ena
\endsubequations
and
\subequations
\bea
H_r({\vec x})&=&N_r({\vec x}) + N_{r-1}({\vec x})-1 \ \ .
\label{4.8a} \\
E_r^+({\vec x})&=&a_r^\dagger({\vec x})
\,a_{r-1}^\dagger({\vec x}) \ \ ,
\label{4.8b} \\
E_r^-({\vec x})&=&{\tilde a}_{r-1}({\vec x})
\,{\tilde a}_r({\vec x}) \ \ ,
\label{4.8c}
\ena
\endsubequations

Using the definitions (\ref{2.63}) and (\ref{2.67}) of
$a({\vec x})$ and ${\tilde a}({\vec x})$ with $\rho_0=\shalf$
and their braiding properties
\footnote{Notice that different
kinds of anyons anticommute with each other,
{\it e.g.} $a_i({\vec x})~a_j({\vec y})+
a_j({\vec y})~a_i({\vec x})=0$ for
$i\not =j$.}, it is possible to check that
\EQ
H_i =\sum_{{\vec x}\in \Omega} H_i({\vec x})
{}~~~~,~~~~
E_i^{\pm} =
\sum_{{\vec x}\in \Omega}E_i^{\pm}({\vec x})
\label{4.0'}
\EN
satisfy the generalized commutation relations (3.1) of the
quantum algebras ${\cal U}_q(A_r)$, ${\cal U}_q(B_r)$
and ${\cal U}_q(D_r)$ with $q= e^{{\rm i}\,\pi\,\nu}$.
A faster way of proving this is to recognize that, after
substituting the definitions (\ref{2.63}) and (\ref{2.67})
of $a({\vec x})$ and
${\tilde a}({\vec x})$ into eqs. (\ref{4.6a}), (\ref{4.7a})
and (\ref{4.8a}),
the local anyonic
generators read as follows
\subequations
\bea
H_i({\vec x})&=&h_i({\vec x})\ \ ,
\label{4.9a} \\
E_i^{+}({\vec x})&=&e_i^{+}({\vec x})
{}~\prod_{{\vec y}\not={\vec x}}
{q^{}_i}^{-{1\over\pi} \Theta({\vec x},
{\vec y})\,h_i({\vec y})}\ \ ,
\label{4.9b} \\
E_i^{-}({\vec x})&=&e_i^{-}({\vec x})
{}~ \prod_{{\vec y}\not={\vec x}}
{q^{}_i}^{{1\over\pi} {\tilde\Theta}({\vec x},
{\vec y})\,h_i({\vec y})}\ \ ,
\label{4.9c}
\ena
\endsubequations
where $h_i({\vec x})$ and $e_i^{\pm}({\vec x})$ are the
generators (\ref{4.1a}), (\ref{4.2a})
and (\ref{4.3a}) in the fundamental
representation satisfying both the
undeformed and the deformed commutation relations.

Expression (\ref{4.9a}) for $H_i({\vec x})$ coincides with
the one obtained with the coproduct.
The same is true also for the operators $E_i^{\pm}({\vec x})$
of eqs. (\ref{4.9b}) and (\ref{4.9c}) if
a suitable gauge is chosen. In fact,
as mentioned in Section 2, the functions
$\Theta({\vec x},{\vec y})$ and ${\tilde\Theta}({\vec x},{\vec y})
=\Theta({\vec y},{\vec x})$,
can be continuously deformed by
gauge transformations keeping the conditions
(\ref{2.a}) and (\ref{2.66'}).
One can therefore choose
\subequations
\EQ
\Theta({\vec x},{\vec y})=\pm \frac{\pi}{2} ~~~~~~~~~~~{\rm for}~~~~
{\vec x} {\scriptstyle{> \atop <}}{\vec y}\ \ ,
\label{4.9'}
\EN
and
\EQ
{\tilde\Theta}({\vec x},{\vec y})=\mp \frac{\pi}{2} ~~~~~~~~~~~
{\rm for}~~~~
{\vec x} {\scriptstyle{> \atop <}}{\vec y}\ \ .
\label{4.9''}
\EN
\endsubequations
Then, in this gauge eqs. (\ref{4.9b})
and (\ref{4.9c}) coincide with eqs. (3.7),
and thus it is clear that the global generators
$H_i$ and $E_i^{+}$ obtained from the anyonic densities (4.9)
satisfy the commutation relations of the deformed
algebras ${\cal U}_q(A_r)$, ${\cal U}_q(B_r)$ and ${\cal U}_q(D_r)$.

Even if we have chosen a particular gauge, we want to stress
that our result is gauge independent because the
braiding relations among anyons and the generalized commutation relations
among
$H_i$ and $E_i^{\pm}$ depend only on the differences
$(\Theta({\vec x},{\vec y}) - \Theta({\vec y},{\vec x}))$ and
$({\tilde\Theta}({\vec x},{\vec y})
- {\tilde\Theta}({\vec y},{\vec x}))$,
and not on the
special form of the functions $\Theta({\vec x},{\vec y})$ and
${\tilde\Theta}({\vec x},{\vec y})$.

\vspace {1cm}

%
%

\sect{Anyonic Construction of ${\cal U}_q(C_r)$}
\vspace{0.7cm}

The anyonic realization of ${\cal U}_q(C_r)$ deserves
a special attention because the
Schwinger construction
of $C_r$ is more natural
in terms of bosonic oscillators and thus involves
all representations. On the contrary, the
previous discussion shows that our realization of
a deformed Lie algebra requires to start from a representation of the
undeformed algebra belonging to the set ${\Re}_{(0,1/2)}$, which is
directly provided by the fermionic Schwinger construction.

To realize the algebra $C_r$ in terms of
fermions, we embed it
into the algebra $A_{2r-1}$ \cite{GNOS}, and introduce
$2r$ fermionic oscillators
$c_{\alpha}({\vec x})~$ $(\alpha=1,...,2r)$
for each point ${\vec x}\in\Omega$.
Then, the generators associated to the short roots $\alpha_i$ of $C_r$
are
\subequations
\bea
h_i({\vec x}) &=&
n_i({\vec x})-n_{i+1}({\vec x})
+n_{2r-i}({\vec x})-n_{2r-i+1}({\vec x})\ \ ,
\label{5.1a}  \\
e_i^+({\vec x})&=&
c_i^\dagger({\vec x})\,c_{i+1}({\vec x})
+c_{2r-i}^\dagger({\vec x})\,c_{2r-i+1}({\vec x})\ \ ,
\label{5.1b} \\
e_i^-({\vec x})&=&
c_{i+1}^\dagger({\vec x})\,c_{i}({\vec x})
+c_{2r-i+1}^\dagger({\vec x})\,c_{2r-i}({\vec x})
\label{5.1c}
\ena
\endsubequations
for $i=1,...,r-1$, while the generators corresponding
to the long root $\alpha_r$ of $C_r$ are
\subequations
\bea
h_r({\vec x})&=&n_{r}({\vec x})-
n_{r+1}({\vec x})\ \ ,
\label{5.2a} \\
e_r^+({\vec x})&=&c_{r}^\dagger({\vec x})
\,c_{r+1}({\vec x})\ \ ,
\label{5.2b} \\
e_r^-({\vec x}) &=&c_{r+1}^\dagger({\vec x})
\,c_{r}({\vec x})\ \ .
\label{5.2c}
\ena
\endsubequations
It is easy
to check that the operators $h_i({\vec x})$,
$e_i^\pm({\vec x})$ defined in these equations
satisfy the commutation relations
(3.5) with the Cartan matrix appropriate for
$C_r$ (see Tab.1) and $q=1$.

Actually, in order to select the fundamental representation we have to
impose a further condition on the
fermionic operators $c_{\alpha}({\vec x})$, namely we must perform a
sort of Gutzwiller projection to force
the fermions to satisfy the extra condition
\[
c_\alpha({\vec x})\,c_\beta({\vec x})=
c_\alpha^\dagger({\vec x})\,c_\beta^\dagger({\vec x}) ~ = ~ 0
\label{5.3}
\]
for any $\alpha,\beta=1,..., 2r$. In this way, the eigenvalues
of $h_i({\vec x})$ are
only $0,\pm1$, and
$(e_i^{\pm}({\vec x}))^2=0$ for $i=1,...,r$. Once this is done,
the representation given by (5.1) and (5.2) belongs to the class
${\Re}_{(0,1/2)}$.

We now observe that to obtain ${\cal U}_q(C_r)$ we cannot simply
replace in eqs.
(5.1) the fermionic oscillators with anyonic
ones defined as we defined them in (\ref{2.63}) and (\ref{2.67}).
In fact, the operators $E^\pm_i({\vec x})$ for $i\not =r$
constructed in this way
could not have the form of eqs. (\ref{4.9b}) and (\ref{4.9c}) because
the disorder
operators in $a_i^\dagger\, a_{i+1}$ would
yield a different structure from the one contained
in $a_{2r-i}^\dagger \,a_{2r-i+1}$.
This difficulty is simply overcome if we require that
the anyons $a_i$ and
$a_{2r-i+1}$ arise from the
fermions $c_i$ and
$c_{2r-i+1}$
coupled both to the same Chern-Simons field with
opposite charges.
Therefore the disorder operators to be used in the Jordan-Wigner
transformation (\ref{2.63}) are
\EQ
K_i({\vec x}) =
K_{2r-i+1}^\dagger({\vec x}) =
{\rm \exp}{\Big[{\rm i} \,
\nu\sum\limits_{{\vec y}\not = {\vec x}}
\Theta({\vec x},{\vec y})\,
    \big(n_i({\vec y})- n_{2r-i+1}({\vec y})\big)}\Big] \ \ .
\label{5.4}
\EN
The same procedure has to be applied also to the oscillators
${\tilde a}_i$ and ${\tilde a}_{2r-i+1}$
by changing $\Theta({\vec x},{\vec y})$ with
${\tilde\Theta}({\vec x},{\vec y})$
in (\ref{5.4}). The anyonic oscillators
defined in this way have the same generalized
commutation relations discussed in Section 2,
and also non trivial
braiding relations among themselves.
For instance, we have
\[
a_i({\vec x})\,a_{2r-i+1}({\vec y})
+ q~a_{2r-i+1}({\vec y})\,a_i({\vec x}) = 0
\, ~~~~~{\rm for}~~~{\vec x}>{\vec y}\ \  .
\]

Now we can replace the fermions with the anyons defined in
this way into eqs. (5.1) and (5.2), and get
\subequations
\bea
H_i~({\vec x})&=&
N_i({\vec x})-N_{i+1}({\vec x})+
N_{2r-i}({\vec x})-N_{2r-i+1}({\vec x})\ \ ,
\label{5.5a} \\
E_i^+({\vec x})&=&
a_i^\dagger({\vec x})\,a_{i+1}({\vec x})+
a_{2r-i}^\dagger({\vec x})\,
a_{2r-i+1}({\vec x})\ \ ,
\label{5.5b} \\
E_i^-({\vec x})&=&
\tilde a_{i+1}^\dagger({\vec x})\,\tilde a_{i}({\vec x})
+\tilde a_{2r-i+1}^\dagger({\vec x})\,
\tilde a_{2r-i}({\vec x})\ \ ,
\label{5.5c}
\ena
\endsubequations
for $i=1,...,r-1$; and
\subequations
\bea
H_r~({\vec x})&=&
N_r({\vec x})-N_{r+1}({\vec x})\ \ ,
\label{5.6a} \\
E_r^+({\vec x})&=&
a_r^\dagger({\vec x})\,a_{r+1}({\vec x})\ \ ,
\label{5.6b}  \\
E_r^-({\vec x})&=&
\tilde a_{r+1}^\dagger({\vec x})\,\tilde a_{r}({\vec x})\ \ .
\label{5.6c}
\ena
\endsubequations
It is immediate to
check that
eqs. (4.9) are reproduced
with $q_r=q^2={\rm e}^{2{\rm i}\pi\nu}$
for the long root, and $q_i=q$ for the
short roots.
Then, the discussion of Section 4 guarantees
that the operators
\[
H_i=\sum_{{\vec x}\in \Omega} H_i({\vec x})~~~~,
{}~~~~E_i^\pm=\sum_{{\vec x}\in \Omega} E_i^\pm({\vec x})
\]
for $i=1,...,r$ satisfy the generalized commutation relations
of the deformed algebra ${\cal U}_q(C_r)$.

\vspace {1cm}

%
%

\sect{Final Remarks}
\vspace{0.7cm}

We have shown that it is possible to establish an explicit
relation between anyons and deformed Lie algebras.
Our treatment has been limited to the anyonic
realization of ${\cal U}_q(G)$ for any classical Lie algebra $G$.
In fact, for our construction it is crucial the existence of
non trivial representations of ${\cal U}_q(G)$ that do not depend
on the deformation parameter $q$ and are therefore common also to
$G$. These representations form a class that we called $\Re_{(0,1/2)}$.
In the case
of classical Lie algebras these representations coincide with the
fundamental ones. This property is shared
also by the exceptional algebras
$E_6$ and $E_7$, and thus we believe that also ${\cal U}_q(E_6)$
and ${\cal U}_q(E_7)$ can be realized in terms of anyons, possibly
by introducing a larger number of them.

The situation is instead quite different for the remaining
exceptional cases
${\cal U}_q(E_8)$,
${\cal U}_q(F_4)$ and ${\cal U}_q(G_2)$,
because the fundamental representations of $E_8$, $F_4$ and $G_2$
are not in the class ${\Re}_{(0,1/2)}$.
Therefore these deformed algebras have no representations
independent from $q$ and this is in contrast
with the possibility of building their
anyonic realization in the way we have discussed.

A second remark is that our construction naturally makes sense
only on a two-dimensional lattice.
However, one can envisage an extension to the case in which anyons are
defined on a continuum two-dimensional space instead of a lattice.
In such a case, one should replace all discrete sums with suitably
defined integrals both in the disorder operator and, more generally,
in the definition of the comultiplication. At present,
this is still an open problem.

On the contrary, it is very easy to ``reduce'' our procedure
to one dimensional chains.
On a chain the ordering is natural and the $q$-commutation
relations like (\ref{2.64}) can be postulated ${\it a priori}$, defining
one-dimensional ``local anyons''.
This amounts simply to replace everywhere the angles
$\Theta({\vec x},{\vec y})$ and
${\tilde\Theta}({\vec x},{\vec y})$ with $\pm{\pi\over 2}$ as specified
in (\ref{4.9'}-\ref{4.9''}).

In such a case it is also possible to assign real values
to the deformation parameter $q$, as in one dimension it
is no longer forced to be a pure phase.
Our construction is valid also in that case. In fact for
real $q$ all our equations still hold, provided that
the creation operators $a_i^\dagger({\vec x})$ and
${\tilde a}_i^\dagger({\vec x})$ are exchanged with each other,
leaving unchanged the destruction operators
$a_i({\vec x})$ and ${\tilde a}_i({\vec x})$.
The case of real $q$ can be interesting because it leads
to unitary representations.

Finally, it is also interesting to notice that our
construction has been extended in \cite{MM}
to the two-parameter deformed Lie algebra $S\ell_{q,s}(2)$ where
the new parameter $s$ is introduced by rotating the reference axes
for the angles $\Theta({\vec x},{\vec y})$
and ${\tilde \Theta}({\vec x},{\vec y})$
in such a way that for generic $s$ ${\tilde \Theta}({\vec x},{\vec y})
\not =\Theta({\vec y},{\vec x})$.

\newpage
\centerline{\bf Figure Captions}
\begin{enumerate}
\item{Exchanging trajectories for anyonic fields.
In Fig. 1a, the anyon in $\vec y$
must move counterclockwise around the anyon in $\vec x$ in order not to
cross its cut. On the contrary in Fig. 1b, the anyon in $\vec y$
must move clockwise around the anyon in $\vec x$ in order not to
cross its cut. Therefore, the orientation of the
exchange trajectories of two
anyons depends on their relative positions.}
\item{Examples of the cuts $\gamma$ for a few points on the lattice.
The dashed
lines represent the reference axes from which the angles
$\Theta$ are measured.
Both the cuts and the
reference axes are on a suitably defined dual lattice
\cite{LS}.}
\item{Examples of the cuts $\delta$ for a few points on the lattice.
The dashed lines represent the reference axes from which the angles
${\tilde \Theta}$ are measured.}
\item{Exchanging trajectories for anyons defined on the lattice with the
angle ${\tilde \Theta}$. They are oriented
in the opposite way as compared with those in Fig. 1.}
\item{Highest weights of the fundamental representations in the
Dynkin bases for the classical Lie algebras and their dimensions.}
\end{enumerate}

\end{document}